\documentclass{PoS}

\title{Array configuration studies for the SKA -- Implementation of figures of merit based on SDR}

\ShortTitle{SKADS: Array Configuration Studies}

\author{\speaker{Dharam V. Lal}%
         \thanks{This effort/activity is supported by the
European Community Framework Programme 6,
Square Kilometer Array Design Studies (SKADS), contract no 011938.
DVL thanks Maxim A. Voronkov for helping with the simulator
in aips$++$ and the related scripts,
and Borkowski, K. \& Fukushima, T. for sharing their
"Cartesian to geodetic coordinates transform" subroutines.}\\
        Max-Planck-Institut f\"ur Radioastronomie\\
        Auf dem H\"ugel 69, 53121 Bonn, Germany\\
        E-mail: \email{dharam@mpifr-bonn.mpg.de}}

\author{Andrei P. Lobanov\\
        Max-Planck-Institut f\"ur Radioastronomie\\
        Auf dem H\"ugel 69, 53121 Bonn, Germany\\
        E-mail: \email{alobanov@mpifr-bonn.mpg.de}}

\abstract{The Square Kilometre Array is going to become operational
at the time when several new large optical, X-ray and Gamma-ray
telescopes are expected to be working.  The main drive for building
the SKA is a significant improvement of sensitivity that would
widen the general scope of the centimetre-wavelength radio science.
To this end, a thorough design studies should be made,
in order to ensure that the SKA becomes a competitive counterpart
of the facilities at other wavebands.

To quantify imaging performance of the SKA configurations,
we are implementing figures of merit based on spatial dynamic range
(SDR).  This work is focused on extensive numerical tests of the analytical,
SDR--based figures of merit derived in the SKA Memo 38 (A. Lobanov).
Here, we present our preliminary results.}

\FullConference{From Planets to Dark Energy: the Modern Radio Universe\\
                 October 1-5 2007\\
                 The University of Manchester, UK}

\begin{document}

\section{Key Terminologies}
\begin{enumerate}
\item {\bf Spatial Dynamic Range}
The ratio of the largest adequately imaged structure
and the synthesized beam.
It depends on the correlator integration time, channel width,
and ($u,v$)-plane coverage. \\ [-0.8cm]
\item {\bf Figures of Merit} Any parameter which is a measure of
($u,v$)-plane coverage;
e.g., spatial dynamic range (SDR), RMS noise levels,
synthesized beam size, etc. \\ [-0.8cm]
\item {\bf $uv$-gap parameter} or {\bf $\Delta u/u$} A
measure of the quality of the ($u,v$)-plane coverage,
characterising the relative size of "holes" in the Fourier plane.
In the simplest case of a circular ($u,v$)-plane coverage,
the $uv$-gap parameter,
$$
\frac{\Delta u}{u} = \frac{(u_2 - u_1)}{u_2},
$$
where $u_1$, $u_2$ ($u_2>u_1$) are the $uv$-radii of
two adjacent baselines. \\ [-0.8cm]
\end{enumerate}

\section{Description of the Work Methodology}

\begin{enumerate}
\item We use the simulator in aips$++$ (ver 1.9 (build 1556))
package to generate a visibility data set for an input array configuration
to determine and study the figures of merit for quantifying
the properties of ($u,v$)-plane coverage and its effect on imaging. \\ [-0.8cm]
\item Next, we vary the array configuration, generate a new visibility data set, and
repeat this exercise many times to complete the full parameter space of SDR.  \\ [-0.8cm]
\item Several scripts, both in Fortran and glish languages, have been written
to accomplish our goal.
We also make use of "classic AIPS" for our imaging analysis. \\ [-0.8cm]
\end{enumerate}

\section{Robustness Checks}

The array can be placed at any position on the Earth's surface,
 but since, we understand the behaviour of the baseline vector for the array
placed at the Earth's pole, when a source being observed is at
90 deg declination, we use this case to understand the
robustness of the output from the aips$++$ simulator.

We generate spiral arrays in many different ways.
We use these array configurations and perform identical pipeline analysis
to conduct robustness checks.

\section{Determination of the "Figures of Merit"}

\paragraph{Experiment 1}

We use equiangular (logarithmic) spiral array configuration, consisting
of a station at the origin, three spiral arms and five stations in each arm,
as the geometry.  The range of baselines used are between
20 to 100~m and 20 to 5000 m.

We make dirty maps (with each image being 4096 pixels $\times$ 4096 pixels
and every pixel being 2~arcsec) for each simulated visibility data set, and
use these maps to determine map characteristics, such as,
peak surface brightness, noise level and beam area.
We use these results
to study the figures of merit
for quantifying the properties of ($u,v$)-plane coverage.

\paragraph{Experiment 2}

Next, in order to probe finer $uv$-gap parameters,
we next perform simulations keeping
largest baseline length, B$_{\rm max}$ (= 5~km) constant and
instead vary $N$; {\i.e.}, we populate a large number of antennas
on to a single equiangular spiral arm pattern in a logarithmic fashion
In addition, as we add
more number of stations for a new array configuration, we increase the
integration time, thereby, we keep identical sensitivities for each new
visibility data set.

Thus, we generate a range of visibility data sets to probe
$uv$-gap parameter from 0.45 ($N$ = 50) to
0.01 ($N$ = 640).
We make dirty and CLEAN maps
(with each image being 8192 pixels $\times$ 8192
pixels and every pixel being 3~arcsec) for each simulated visibility data set,
and use the CLEAN maps to determine several map characteristics.

\paragraph{Simulations: Parameters Used}

We use the following parameters for the generation of a range of visibility
data sets: \\ [-0.2cm]

\noindent
\begin{tabular}{llll||lllll}
\multicolumn{3}{c}{Telescope settings} & & & \multicolumn{4}{c}{Input group of
source components} \\
\hline
& & & & & & & & \\
Frequency &:&  L band (1.4 GHz)            & & &    & \multicolumn{3}{l}{Experiment~1} \\
Antenna &:&  25 m (VLA)                    & & &    &\multicolumn{3}{l}{(\hbox{Constant $N$ and Varying B$_{\rm max}$})} \\
Bandwidth &:&  3.2 MHz                     & & & \multicolumn{4}{l}{Six Gaussian components;} \\
No. of channels & :  &  1                  & & & \multicolumn{4}{r}{typical source size, $\sim$1$^{\prime}$} \\
Direction (J2000) &:& 00:00:00 $+$90.00.00 & & & \multicolumn{4}{c}{-----------------------------------------------} \\
Elevation limit &: & 12 deg                & & &    & \multicolumn{3}{l}{Experiment~2} \\
Shadow limit &: &   0.001                  & & &    & \multicolumn{3}{l}{(\hbox{Constant B$_{\rm max}$ and Varying $N$})} \\
Start\_time (IAT) &:& 22/08/2007 / 06:00   & & &  \multicolumn{4}{l}{Seven Gaussian components;} \\
Stop\_time (IAT) &:&  22/08/2007 / 18:00   & & &  \multicolumn{4}{r}{source size range: $0.1^{\prime} \times 0.1^{\prime} - 120^{\prime} \times 40^{\prime}$}
\end{tabular}

\section{Preliminary Inferences}

\begin{enumerate}
\item[$\Rightarrow$] The behaviour of figures of merit and hence
the SDR does not seem to have a simple dependence on~$\Delta u / u$.
Our results show that close to small $uv$-gap parameter values,
the (nearly) linear relationship does not hold good. \\ [-0.8cm]
\item[$\Rightarrow$] Our simulations show that the $uv$-gap parameter
can be used to relate the ($u,v$)-plane coverage to the characteristics
of the map.
Although, we cover a small part of the full parameter space to
be probed, the preliminary study demonstrates a valuable result which is
important to put constraints on the SKA design studies. \\ [-0.8cm]
\end{enumerate}

\section{Conclusions}

To make the SKA a competitive instrument that would match the
capabilities of future optical and X-ray telescopes, two basic
conditions must be fullfilled: (i) resolution $\lesssim$ 1 mas at
the high end of frequency and (ii) Fourier plane filling factor
$\Delta u / u$ $\lesssim$ 0.05 over entire range of ($u,v$)-plane coverage.

Any compromise in either of these two conditions would reduce the
imaging capability and narrow the scientific scope of the SKA.


\end{document}